\documentclass[prd,twocolumn,showpacs,english]{revtex4}
\usepackage{amsmath}
\usepackage{amsfonts}
\usepackage{amssymb}
\usepackage{color}
\usepackage[unicode=true,pdfusetitle,
 bookmarks=true,bookmarksnumbered=true,bookmarksopen=true,bookmarksopenlevel=3,
 breaklinks=false,pdfborder={0 0 1},backref=false,colorlinks=true]
 {hyperref}

\usepackage{enumitem}

\newcommand{\udt}[3]{#1^{#2}_{\phantom{#2}#3}}

\newcommand{\dut}[3]{#1_{#2}^{\phantom{#2}#3}}

\begin{document}
\title{Polarization of gravitational waves in symmetric teleparallel theories
of gravity and their modifications}
\author{Ismail Soudi}
\email{ismail.soudi@gmail.com}

\affiliation{Institute of Space Sciences and Astronomy, University of Malta, Msida,
MSD 2080, Malta}
\affiliation{Department of Physics, University of Malta, Msida, MSD 2080, Malta}
\author{Gabriel Farrugia}
\email{gabriel.farrugia.11@um.edu.mt}

\affiliation{Institute of Space Sciences and Astronomy, University of Malta, Msida,
MSD 2080, Malta}
\affiliation{Department of Physics, University of Malta, Msida, MSD 2080, Malta}
\author{Viktor Gakis}
\email{vgakis@central.ntua.gr}

\affiliation{Institute of Space Sciences and Astronomy, University of Malta, Msida,
MSD 2080, Malta}
\affiliation{Department of Physics, National Technical University of Athens, Zografou
Campus GR 157 73, Athens, Greece}
\author{Jackson Levi Said}
\email{jackson.said@um.edu.mt}

\affiliation{Institute of Space Sciences and Astronomy, University of Malta, Msida,
MSD 2080, Malta}
\affiliation{Department of Physics, University of Malta, Msida, MSD 2080, Malta}
\author{Emmanuel N. Saridakis}
\email{Emmanuel_Saridakis@baylor.edu}

\affiliation{Department of Physics, National Technical University of Athens, Zografou
Campus GR 157 73, Athens, Greece}
\affiliation{CASPER, Physics Department, Baylor University, Waco, TX 76798-7310,
USA}


\begin{abstract}
Symmetric teleparallel gravity (STG) offers an interesting avenue to 
formulate a theory of gravitation that relies neither on curvature nor torsion but 
only on non-metricity $Q$. Given the growing number of confirmed observations of gravitational 
waves (GWs) and their use to explore gravitational theories, in this 
work we investigate the GWs in various extensions of STG, focusing on their speed and 
polarization. We explore the plethora of theories that this new framework opens up, that is, as general relativity (GR) can be modified, so to can the symmetric teleparallel equivalent of general relativity (STEGR). In this work, we investigate the fate of GWs in the generalized irreducible decomposition of STEGR, generalizations of the STEGR Lagrangian, $f(Q)$, a scalar field nonminimally coupled to the STEGR Lagrangian, and the general setup of $f(Q,B)$ theory where $B$ is the boundary term difference between the Ricci scalar and the STEGR Lagrangian. Coincidentally, $f(Q,B)$ forms a more general theory than $f(R)$ gravity since $Q$ embodies the second order elements of the Ricci scalar while $B$ takes on it's fourth order boundary terms. Our work deals mainly with the resulting scalar-vector-tensor polarization modes of the plethora of STG theories, and how they effect their respective speeds of propagation.
\end{abstract}
\maketitle

\section{Introduction}

To adequately describe the late-time behavior of the Universe and
the behavior of galactic dynamics, on top of Einstein's theory of
general relativity (GR) \cite{carroll2004spacetime} one necessitates
the introduction of the so-called dark matter and dark energy 
sectors \cite{dodelson2003modern}, resulting to the $\Lambda$CDM paradigm. This stems from
the fact that the Universe is observed to be accelerating in its expansion
\cite{Riess:1998cb,Riess:2004nr} and that galaxies do not contain
enough matter to sustain their measured rotational curves \cite{Rubin:1970zza}.
While recent Planck Collaboration results shows mild tension in this picture
of the Universe \cite{Aghanim:2018eyx}, the theory hits its breaking
point when the early Universe is investigated. This regime of exploration
has led to the suggestion of a scalar field to explain the epoch of
inflation \cite{Guth:1980zm,Linde:1981mu} which may solve some other
problems too \cite{Weinberg:1988cp}. Given the large body of
research investigating new physics at early-times, we are motivated to
explore alternative theories of gravity in other regimes, namely in
the strong field regime where gravitational wave (GW) radiation is
emitted.

Modifications to GR mainly come in the form of extensions to the Einstein-Hilbert
action, namely raising the Ricci scalar $\mathcal{R}$ to an arbitrary
function Lagrangian $f(\mathcal{R})$ \cite{DeFelice:2010aj}, where
$\mathcal{R}$ is determined using the Levi-Civita connection (this character style is used throughout to denote those quantities calculated using the the Levi-Civita connection). Extended models of gravity
\cite{Capozziello:2011et} also include other scalar invariants that
contribute to the Lagrangian, such as the Gauss-Bonnet term in 
$f(\mathcal{R},\mathcal{G})$
gravity \cite{Bamba:2009uf,Elizalde:2010jx} and may even include
a nonminimal coupling with the trace of the stress-energy tensor through
$f(\mathcal{R},T)$ gravity \cite{Alves:2016iks,Harko:2011kv,Alvarenga:2013syu}. While 
many potential extensions to GR exist,
alternatives to GR are more difficult to be constructed and require
more intense exploration due to their fundamental reconstitution of basic tenants. An interesting class of alternative theories of
gravity comes with the use of torsion instead of curvature, i.e start
from the teleparallel equivalent of general relativity \cite{Pereira.book}
and construct modifications like $f(T)$ gravity \cite{Cai:2015emx},
$f(T,T_{G})$ gravity \cite{Kofinas:2014owa,Kofinas:2014daa}, $f(T,\mathcal{T})$
gravity \cite{Harko:2014aja} etc. Some other examples include massive
gravity where gravitational waves (GWs) are endowed with a nonvanishing
effective mass \cite{deRham:2010kj}, while Ho\v{r}ava-Lifshitz gravity
reexamines the relation between space and time in the quantum regime
\cite{Horava:2009uw}. As a shared goal, all these modifications aim
to confront observations at all scalaes better than $\Lambda$CDM \cite{Clifton:2011jh,Cai:2009zp}.

Symmetric teleparallel gravity (STG) is an interesting theory in that
it can describe gravitation while retaining a vanishing contribution
of both curvature and torsion, which geometrically implies that vectors do
remain parallel at long distances on a manifold \cite{Nester:1998mp}.
In this scenario, gravity is manifested through a nonmetricity scalar
$Q$ that gives a measure of the amount of nonmetricity present given a particular metric tensor solution which is analogous to the Ricci scalar in GR
\cite{BeltranJimenez:2017tkd,Conroy:2017yln,Golovnev:2017dox}.
In terms of the affine connection of the general frames, by demanding that the curvature
vanishes and that the connection is torsionless (symmetric indices)
then the remaining gravitational information will be encoded in nonmetricity
contributions \cite{Jimenez:2015fva}. The latter assumption of vanishing
torsion can also lead to its own version of gravitation. In STG the
metricity condition of GR is relaxed resulting in the symmetric teleparallel gravity equivalent of general relativity (STEGR), which is analogous to the procedure where the vanishing torsion condition leads directly to the teleparallel equivalent of general relativity (TEGR) scalar \cite{Cai:2015emx}. Teleparallel gravity and STG share a number of important properties, one of which is their ability to separate gravitational and inertial effects \cite{Koivisto:2018aip} which is not possible in GR. This has resulted in many strains on GR theory such as the issue of defining a gravitational energy-momentum tensor \cite{Brown:1992br}. STG can also be shown to be consistent with a connection that simplifies to a partial derivative through the so-called coincident gauge \cite{BeltranJimenez:2017tkd,Harko:2018gxr}.

By construction, the nonmetricity scalar, $Q$, is equivalent to the Ricci
scalar up to a boundary term in the Lagrangian \cite{Runkla:2018xrv}.
This takes the form of 
\begin{equation}
\mathcal{R}=Q-B,\label{ricci_nonme_equiv}
\end{equation}
where $B$ is the boundary term. This means that such a Lagrangian
would agree with GR at the level of field equations at all levels
of the classical regime, and so produce no measurable differences \cite{Harko:2018gxr}.
As with the $f(R)$ extensions \cite{DeFelice:2010aj}, one can construct a plethora of extended STG theories. On the face of it, the relation in Eq.(\ref{ricci_nonme_equiv} can be
interpreted as a breakdown of the Ricci scalar into second order contributions
and its boundary contribution made by fourth-order parts. In this way, even
$f(\mathcal{R})$ theory can be made richer by raising it to the broader
class of $f(Q,B)$ theories, where both class of contributions can be fixed independently.

Straightforward modifications of the STEGR Lagrangian gives directly  $f(Q)$ gravity, which has the advantageous property that the field equations remain
second order irrespective of the particular Lagrangian function. This is in contrast to $f(\mathcal{R})$ gravity scenario \cite{Harko:2018gxr} where all nontrivial models lead to fourth order field equation.
The result of this observation is that despite the linear cases being
equal up to a boundary term (as shown in (\ref{ricci_nonme_equiv})),
the general scenarios of arbitrary functions are not equivalent. This
inequality stems from the boundary term which no longer remains linear
(i.e. a boundary term) in the generalized case, namely
\begin{equation}
f(\mathcal{R})\neq f(Q).
\end{equation}
These theories are only equal when the argument of the arbitrary Lagrangian takes on the exact form $f(Q,B)=f(Q-B)=f(\mathcal{R})$, where the STG Lagrangian is complemented by a boundary term argument as well.

In curvature based gravity theory it is well-known that one can add a scalar field nonminimally
coupled to the Ricci scalar $\mathcal{R})$ 
\cite{Sahni:1998at,Faraoni:2000wk,Fujii:2003pa},
and similarly in teleparallel gravity one can add a scalar field nonminimally
coupled to the torsion scalar $T$ \cite{Geng:2011aj,Xu:2012jf,Hohmann:2018rwf}.
In Ref.\cite{Harko:2018gxr}, the possibility of nonminimally coupled
general function of the nonmetricity scalar is considered with interesting
results. However, not much work has been done on other scalar invariant
generalizations such as Gauss-Bonnet extensions. The possibility of
a scalar field coupled to STG has been explored in a number of recent
works \cite{Jarv:2018bgs,Runkla:2018xrv} where the nonminimal coupling case was investigated. This is an interesting possibility for the extended
$f(Q,B)$ context due to the separation between second and fourth
order contributions.

STG also offers another interesting way to investigate
gravitational models, since the nonmetricity scalar equivalent of GR can
be separated into five irreducible components 
\cite{BeltranJimenez:2017tkd,BeltranJimenez:2018vdo}.
These irreducibles can then be generalized linearly to form a completely
new avenue for gravitational modification. While the $f(Q,B)$ scenario
has a clear $f(\mathcal{R})$ limit, the generalized irreducible context
is only fixed by the GR scenario, since the boundary term is not necessarily
included in that form of the theory.

On the other hand, observations of gravitational radiation have confirmed
not only the existence of GWs as the mediator of gravitational information
\cite{Abbott:2016blz} but also opened the possibility of setting bounds on the possible polarization modes that a GW event would propagate \cite{Abbott:2017oio}.
This is a crucial component to testing gravitational models due to
its inherently model-independent nature. Beyond this comparison, source
modeling techniques would necessarily have to be employed \cite{Barack:2018yly}.

In Ref.\cite{Hohmann:2018wxu}, GWs polarization modes are investigated
for the general linear case of nonmetricity scalar irreducible components, using
the Newman-Penrose formalism. In this work, the foundations of exploring
polarization modes in STG have been laid. In the current work, we
are interested in extending this work to further extensions and scenarios
that have appeared in the STG literature and which show promise in
terms of realistic theories of gravity \cite{Heisenberg:2018vsk}.

The paper is divided as follows. In section \ref{modell} the foundations of STG
are introduced, with some discussion on its relation to GR. Section
\ref{secIII} then delves into the potential polarizations of GWs in the generalized
GR equivalent, that is $f(Q)$ gravity. The possibility of nonminimally
coupled scalar fields is advanced in section \ref{secIV}. Section \ref{secV}   
extends the generalized scenario to the $f(Q,B)$ theory, where we
can compare the results of GWs with $f(\mathcal{R})$ gravity
for the choice of $f(Q,B)=f(Q-B)$. Finally the main conclusions are
discussed and summarized in section \ref{Conclusions}. Throughout the work, geometric
units are used unless otherwise stated.

\section{Symmetric Teleparallel Gravity (STG)}
\label{modell}

In this section we present STG from its foundations. We start by remarking that the mechanism by which gravity is mediated
is an expression of the affine connection and not the physical manifold
\cite{Nakahara:2003nw,misner2017gravitation,ortin2007gravity}. For instance, in GR, the property of curvature emerges through the Levi-Civita connection and not the manifold which is described through the metric tensor, and thus the connection can be equally described by other properties such as nonmetricity. By the strong equivalence principle \cite{misner2017gravitation},
every point on the manifold has a well-defined tangent space, where
the connection acts as an intermediary between neighboring tangent
spaces so that derivative operators can be defined. This implies that
the decomposition of a general affine connection can be written as \cite{ortin2007gravity} 
\begin{equation}
\udt{\Gamma}{\alpha}{\mu\nu}=\mathring{\Gamma}_{\mu\nu}^{\alpha}+\udt{K}{\alpha}{\mu\nu}
+\udt{L}{\alpha}{\mu\nu},
\end{equation}
where $\mathring{\Gamma}_{\mu\nu}^{\alpha}$ is the Christoffel symbols
of the Levi-Civita connection, $\udt{K}{\alpha}{\mu\nu}$ is the contorsion
tension representing the difference between the Christoffel symbols
and the teleparallel connection (i.e. the  the Weitzenb\"{o}ck  connection),
and $\udt{L}{\alpha}{\mu\nu}$ is the \textit{disformation} tensor
which encodes the nonmetricity contribution due to the nonmetricity
tensor \cite{Nester:1998mp} 
\begin{equation}
Q_{\alpha\mu\nu}:=\nabla_{\alpha}g_{\mu\nu}.
\end{equation}
The disformation takes the explicit form of \cite{BeltranJimenez:2017tkd}
\begin{equation}
\udt{L}{\alpha}{\mu\nu}=\frac{1}{2}g^{\alpha\beta}\left(-Q_{\mu\beta\nu}-Q_{\nu\beta\mu}
+Q_{\beta\mu\nu}\right).
\end{equation}

The Einstein-Hilbert Lagrangian can equivalently be written as
\cite{Harko:2018gxr} 
\begin{equation}
R=\mathcal{L}_{E}+\mathcal{L}_{B},
\end{equation}
where $\mathcal{L}_{E}$ represents Einstein's original Lagrangian
from the Levi-Civita connection \cite{Einstein:1916cd,Harko:2018gxr}
\begin{equation}
\mathcal{L}_{E}:=g^{\mu\nu}\left(\mathring{\Gamma}_{\beta\mu}^{\alpha}\mathring{\Gamma}_{
\nu\alpha}^{\beta}-\mathring{\Gamma}_{\beta\alpha}^{\alpha}\mathring{\Gamma}_{\mu\nu}^{
\beta}\right) ,
\end{equation}
and the total derivative (or boundary term) is given by 
\begin{equation}
\mathcal{L}_{B}=g^{\alpha\mu}\mathcal{D}_{\alpha}\mathring{\Gamma}_{\mu\nu}^{\nu}-g^{
\mu\nu}\mathcal{D}_{\alpha}\mathring{\Gamma}_{\mu\nu}^{\alpha},
\end{equation}
where $\mathcal{D}_{\alpha}$ represents the covariant derivative
with respect to the Levi-Civita connection. This higher derivative
version of the equivalent Lagrangian $\mathcal{L}_{E}$  is ubiquitously
adopted due to its covariance, while $\mathcal{L}_{E}$ alone is not
covariant within the Levi-Civita connection setting. STG approaches this issue by promoting the partial derivative
to a covariant operator, called the \textit{coincident gauge}, where
gravitation is no longer mediated through the connection (called the
``Palatini connection'') \cite{Koivisto:2018aip}. The disformation
then takes the form 
\begin{equation}
\udt{L}{\alpha}{\beta\gamma}=-\frac{1}{2}g^{\alpha\lambda}\left(\nabla_{\gamma}g_{
\beta\lambda}+\nabla_{\beta}g_{\lambda\gamma}-\nabla_{\lambda}g_{\beta\gamma}
\right).\label{disformation_chris_equiv}
\end{equation}
By the coincident gauge ($\nabla_{\alpha}\rightarrow\partial_{\alpha}$),
the disformation is essentially the negative of the Christoffel symbols.
The GR equivalent Lagrangian then turns out to simply be \cite{Harko:2018gxr}
\begin{equation}
Q=-g^{\mu\nu}\left(\udt{L}{\alpha}{\beta\mu}\udt{L}{\beta}{\nu\alpha}-\udt{L}{\alpha}{
\beta\alpha}\udt{L}{\beta}{\mu\nu}\right),\label{Q_Scalar}
\end{equation}
which is simply the negative sign of $\mathcal{L}_{E}$, and can equivalently
be interpreted in terms of the Christoffel symbols due to 
Eq.(\ref{disformation_chris_equiv}).
This produces the exact same relations as the Einstein field equations at the level of the field equations.

In order to consider generalizations of the GR formalism in the STG
context we consider the general action \cite{Koivisto:2018aip}
\begin{equation}
S_{G}={\displaystyle \int d^{4}x\,\left[\sqrt{-g}f + \dut{\lambda}{\alpha}{\beta\mu\nu}\udt{R}{\alpha}{\beta\mu\nu}+\dut{\lambda}{\alpha}{\mu\nu}\udt{T}{\alpha}{\mu\nu}\right],}
\end{equation}
where the Lagrangian assumes a Palatini approach with $f=f(g_{\mu\nu},\udt{\Gamma}{\alpha}{\mu\nu})$, and Lagrange multipliers, $\dut{\lambda}{\alpha}{\beta\mu\nu}$ and $\dut{\lambda}{\alpha}{\mu\nu}$, are used to eliminate the curvature-full Riemann tensor and torsion-full torsion tensor.

Straightforwardly, a conjugate to the Lagrangian can be defined as
\cite{BeltranJimenez:2017tkd} 
\begin{equation}
\udt{P}{\alpha}{\mu\nu}:=\frac{1}{2}\frac{\partial 
f}{\partial\dut{Q}{\alpha}{\mu\nu}},
\label{conj_Lag_ten}
\end{equation}
which yields the metric tensor field equations
\begin{equation}
\frac{2}{\sqrt{-g}}\nabla_{\alpha}\left(\sqrt{-g}\udt{P}{\alpha}{\mu\nu}\right)-\frac{
\partial f}{\partial g^{\mu\nu}}-\frac{1}{2}fg_{\mu\nu}=T_{\mu\nu},\label{STG_field_eqn}
\end{equation}
where 
\begin{equation}
T_{\mu\nu}:=-\frac{2}{\sqrt{-g}}\frac{\delta(\sqrt{-g}\mathcal{L}_{m})}{\delta g^{\mu\nu}},
\end{equation}
is the regular energy-momentum tensor for matter. Secondly, the action can also be varied with respect to the connection since the Palatini approach is being adopted here which results in \cite{BeltranJimenez:2018vdo} 
\begin{equation}
\nabla_{\rho}\dut{\lambda}{\alpha}{\nu\mu\rho}+\dut{\lambda}{\alpha}{\mu\nu} = \sqrt{-g} \udt{P}{\mu\nu}{\alpha} + \dut{H}{\alpha}{\mu\nu},
\end{equation}
where
\begin{equation}
\dut{H}{\alpha}{\mu\nu} := -\frac{1}{2}\frac{\delta(\sqrt{-g}\mathcal{L}_{m})}{\delta \udt{\Gamma}{\alpha}{\mu\nu}},
\end{equation}
is the hypermomentum. The Lagrange multipliers can be eliminated through symmetry considerations to give the relation \cite{BeltranJimenez:2018vdo} 
\begin{equation}\label{connection_field_eq}
\nabla_{\mu}\nabla_{\nu}\left(\sqrt{-g}\udt{P}{\mu\nu}{\alpha}\right)=0,
\end{equation}
which can be interpreted as the connection field equations that is trivially solved by the coincident gauge choice. Here it is assumed that $\nabla_{\mu}\nabla_{\nu} \dut{H}{\alpha}{\mu\nu}$ vanishes.

\section{Gravitational waves in extensions of STG}
\label{secIII}

In this section we investigate GWs in two extensions of STG, namely 
the perturbed versions of the generalized irreducible decomposition of the STG which emerges from nonmetricity
scalar $Q$, as well as the other natural generalization of the theory, namely $f(Q)$ gravity.

\subsection{GWs in the generalized irreducible decomposition of STG}
\label{secIIIA}

The most general quadratic scalar built form irreducible components
of the nonmetricity tensor is given by 
\begin{align}
\mathbb{Q} & 
:=c_{1}Q_{\alpha\mu\nu}Q^{\alpha\mu\nu}+c_{2}Q_{\alpha\mu\nu}Q^{\mu\alpha\nu}+c_{3}Q_{
\alpha}Q^{\alpha}\nonumber \\
 &\ \ \ \, +c_{4}\bar{Q}_{\alpha}\bar{Q}^{\alpha}+c_{5}\bar{Q}_{\alpha}Q^{\alpha},
 \label{eq:Gen Q scalar}
\end{align}
where $c_{1},..,c_{5}$ are arbitrary constants. Note that this is quadratic at the level of the Lagrangian, and does not refer to the order of the resulting theory which can be generalized to $f(Q)$ and still kept at second order. To reproduce the
STG equivalent of GR, i.e. the so-called STEGR, a unique choice of these parameters must be considered, and it turns out to be 
\begin{equation}
Q:={\displaystyle \frac{1}{4}}Q_{\alpha\mu\nu}Q^{\alpha\mu\nu}-{\displaystyle 
\frac{1}{2}}Q_{\alpha\mu\nu}Q^{\mu\alpha\nu}-{\displaystyle 
\frac{1}{4}}Q_{\alpha}Q^{\alpha}+{\displaystyle 
\frac{1}{2}}\bar{Q}_{\alpha}Q^{\alpha},\label{eq: Q GR}
\end{equation}
which can be shown to be equal to Eq.(\ref{Q_Scalar}) \cite{BeltranJimenez:2017tkd}. This formulation
of the theory is not possible in GR and can offer an interesting perspective
on generalizing the STG equivalent of GR.

To derive the field equations for this Lagrangian we first need to
determine the conjugate to the Lagrangian, which reads as
\begin{align}
\udt{P}{\alpha}{\mu\nu} &= \frac{1}{2}\Bigg[c_{1}\udt{Q}{\alpha}{\mu\nu}+c_{2}Q_{(\mu}{}^{\alpha}{}_{\nu)}+c_{3}g_{
\mu\nu}Q^{\alpha}\nonumber \\
 & \quad\quad \ \
c_{4}\delta{}^{\alpha}{}_{(\mu}\widetilde{Q}_{\nu)}+\frac{c_{5}}{2}\left(\widetilde{Q}
^{\alpha}g_{\mu\nu}+\delta{}^{\alpha}{}_{(\mu}Q_{\nu)}\right)\Bigg].
\label{eq:P for Q}
\end{align}
For convenience, we define the tensor quantity 
\cite{BeltranJimenez:2018vdo,Heisenberg:2018vsk}
\begin{align}
q_{\mu\nu} & :=2{\displaystyle \frac{\partial\sqrt{-g}f}{\partial 
g^{\mu\nu}}}-\sqrt{-g}fg_{\mu\nu}\nonumber \\
 & 
=\sqrt{-g}\Biggl[c_{1}\left(Q_{\alpha\beta\mu}\udt{Q}{\alpha\beta}{\nu}-Q_{\mu\alpha\beta}\dut{Q}{\nu}{\alpha\beta}\right)\nonumber \\
 & 
\quad+c_{2}Q_{\alpha\beta\mu}Q^{\beta\alpha}{}_{\nu}+c_{3}\left(2Q_{\alpha}\udt{Q}{\alpha}{\mu\nu}-
Q_{\mu}Q_{\nu}\right)\nonumber \\
 & 
\quad+c_{4}\widetilde{Q}_{\mu}\widetilde{Q}_{\nu}+c_{5}\widetilde{Q}_{\alpha}\udt{Q}{\alpha}{\mu\nu}\Biggl],
\end{align}
with the help of which  the field equations take the elegant form 
\begin{equation}
4\nabla_{\alpha}\left(\sqrt{-g}\udt{P}{\alpha}{\mu\nu}\right)-q_{\mu\nu}-\sqrt{-g}fg_{\mu\nu
} = T_{\mu\nu},\label{eq:EQM for f=00003D00003D00003DQ/2 final}
\end{equation}
where vacuum background is already assumed. For our purposes we take the general irreducible decomposition of
the nonmetricity scalar, that is $f(g_{\mu\nu},\udt{\Gamma}{\alpha}{\mu\nu})=\mathbb{Q}$.

We proceed, by perturbing the metric tensor in a Minkowski background setting up to first order so that
\begin{equation}
g_{\mu\nu}=\eta_{\mu\nu}+\epsilon_1 h_{\mu\nu},\label{metric_pert}
\end{equation}
where $\eta_{\mu\nu}$ is the Minkowski metric, $\epsilon_1$ is a first-order parameter, and 
$h_{\mu\nu}$ is the perturbation of $g_{\mu\nu}$
namely $h_{\mu\nu}:=\delta g_{\mu\nu}$. In general, we can perturb
any metric-dependent quantity $A\left(g\right)$ up to first order
through 
\begin{equation}
A\left(\eta_{\mu\nu}+\epsilon_1
h_{\mu\nu}\right)=A\left(\eta_{\mu\nu}\right)+\epsilon_1\delta 
A\left(\eta_{\mu\nu},h_{\mu\nu}\right).
\end{equation}
A subcase of the nonmetricity covariant derivative is the usual partial
derivative in the coincident gauge. We will use interchangeably the
notation $A^{(1)}=\delta A$ to indicate the first order perturbation
of a quantity $A$. We use this method to find the first order part
of the field equations Eq.(\ref{eq:EQM for f=00003D00003D00003DQ/2 final})
\begin{equation}
4\nabla_{\alpha}\delta\left[\left(\sqrt{-g}\udt{P}{\alpha}{\mu\nu}\right)\right]-\delta 
q_{\mu\nu}-\delta\left(\mathcal{L}_{Q}g_{\mu\nu}\right) = \delta T_{\mu\nu}.
\label{eq:=00003D00003D0003B4EQM}
\end{equation}

One would start by calculating all the relevant quantities, $\delta\left(\sqrt{-g}\udt{P}{\alpha}{\mu\nu}\right)$,
$\delta\left(\mathcal{L}_{Q}g_{\mu\nu}\right)$ and $\delta q_{\mu\nu}$, in the coincident gauge. However, it turns out the only nonvanishing quantity at first order is $\delta\left(\sqrt{-g}\udt{P}{\alpha}{\mu\nu}\right)$,
which is not unexpected since the other quantities are third order
in the metric. This gives 
\begin{align}
\delta\left(\sqrt{-g}\udt{P}{\alpha}{\mu\nu}\right) & 
=c_{1}\partial^{\alpha}h_{\mu\nu}+\tfrac{1}{2}c_{2}(\partial_{\mu}\udt{h}{\alpha}{\nu}
+\partial_{\nu}\udt{h}{\alpha}{\mu})\nonumber \\
 & +c_{3}\eta_{\mu\nu}\partial^{\alpha}h\nonumber \\
 & 
+\tfrac{1}{4}c_{4}(2\delta^{\alpha}{}_{\nu}\partial_{\alpha_{1}}h_{\mu}{}^{\alpha_{1}}
+2\delta^{\alpha}{}_{\mu}\partial_{\alpha_{1}}h_{\nu}{}^{\alpha_{1}})\nonumber \\
 & 
+\tfrac{1}{4}c_{5}(2\eta_{\mu\nu}\partial_{\alpha_{1}}h^{\alpha\alpha_{1}}+\delta^{\alpha}
{}_{\nu}\partial_{\mu}h+\delta^{\alpha}{}_{\mu}\partial_{\nu}h),\nonumber \\
\label{eq:=00003D00003D0003B4gP}
\end{align}
and substituting Eq.\ref{eq:=00003D00003D0003B4gP} back to Eq.\ref{eq:=00003D00003D0003B4EQM}, the linearised field equations read as 
\begin{align}
\delta T_{\mu\nu} & =c_{1}\Box h_{\mu\nu}\nonumber \\
 & 
+\tfrac{1}{2}\left(c_{2}+c_{4}\right)(\partial_{\alpha}\partial_{\mu}h_{\nu}{}^{\alpha}
+\partial_{\alpha}\partial_{\nu}h_{\mu}{}^{\alpha})\nonumber \\
 & +c_{3}\eta_{\mu\nu}\partial_{\alpha_{1}}\partial^{\alpha_{1}}h\nonumber \\
 & 
+\tfrac{1}{2}c_{5}(\eta_{\mu\nu}\partial_{\alpha_{1}}\partial_{\alpha}h^{\alpha\alpha_{1}}
+\partial_{\nu}\partial_{\mu}h).
\label{eq:=00003D00003D00003D0003B4E QM semifinal}
\end{align}



Let us stress at this point that in order to study the above linearized field equations, we no longer have the usual diffeomorphism invariance which would have allowed us to use the traceless transverse gauge or any gauge that is sourced from a coordinate change. This is due to the fact that we have already fixed a specific coordinate frame in which our connection
trivializes to the coincident gauge. Therefore to further study Eq.(\ref{eq:=00003D00003D00003D0003B4E QM semifinal}),
we need to proceed in the most general way possible by performing
the full scalar-vector-tensor (SVT) decomposition (for details see on this approach in appendix \ref{subsec:Scalar-Vector-Tensor}).

Inserting Eq.(\ref{eq:hS}) into Eq.(\ref{eq:=00003D00003D00003D0003B4E QM semifinal})
and then Fourier transforming the space part of the perturbation through
$h_{\mu\nu}(x^{i},t)\rightarrow \operatorname{Re}\left(h_{\mu\nu}(t)e^{ik_{i}x^{i}}\right)$ we
obtain the following field equations for the scalar perturbations 
\begin{align}
\delta T^{00} =& -2(6c_{3}+c_{5})k^{2}\psi-6(2c_{3}+c_{5})\ddot{\psi}\label{eq:=0003B4T00scalar}\\
 & +4(c_{1}+c_{3})k^{2}\varphi+4(c_{1}+c_{2}+c_{3}+c_{4}+c_{5})\ddot{\varphi}\nonumber \\
 & -(2c_{3}+c_{5})k^{4}E-(2c_{3}+c_{5})k^{2}\ddot{E}\nonumber \\
 & -2(c_{2}+c_{4}+c_{5})k^{2}\dot{B},\nonumber 
\end{align}
\begin{align}
-i\frac{k_{i}}{k^{2}}\delta T^{0i} =& (2c_{1}+c_{2}+c_{4})(k^{2}B+\ddot{B})\label{eq:=0003B4T0iscalar}\\
 & +(c_{2}+c_{4}+c_{5})(k^{2}\dot{E}-2\dot{\varphi})\nonumber \\
 & +2(c_{2}+c_{4}+3c_{5})\dot{\psi},\nonumber 
\end{align}
\begin{align}
\delta_{ij}\delta T^{ij} =& 12(c_{1}+3c_{3})\ddot{\psi}\label{eq:=0003B4ij=0003B4Tijscalar}\\
 & +4(3c_{1}+c_{2}+9c_{3}+c_{4}+3c_{5})k^{2}\psi\nonumber \\
 & -6(2c_{3}+c_{5})\ddot{\varphi}-2(6c_{3}+c_{5})k^{2}\varphi\nonumber \\
 & +2(c_{1}+c_{2}+3c_{3}+c_{4}+2c_{5})k^{4}E\nonumber \\
 & +2(c_{1}+3c_{3})k^{2}\ddot{E}+2(c_{2}+c_{4}+3c_{5})k^{2}\dot{B},\nonumber 
\end{align}
\begin{align}
\sigma =& 2(c_{2}+c_{4})\dot{B}-2c_{5}\varphi+\bigl(4(c_{2}+c_{4})+6c_{5}\bigr)\psi\label{eq:=0003C3 scalar}\\
 & +\bigl(2(c_{1}+c_{2}+c_{4})+c_{5}\bigr)k^{2}E+2c_{1}\ddot{E},\nonumber 
\end{align}
where $\sigma$ is a scalar that generates a part of the anisotropic
tensor defined by $\delta T^{\mu\nu}$.

From the scalar perturbation equations, one can calculate the dispersion relation
for the scalar modes by Fourier transforming the time derivatives
$A(t)\rightarrow \operatorname{Re}\left(Ae^{-i\omega t}\right)$ and then evaluating the equations
in vacuum, which results in
\begin{align}
c_{1}\kappa_{1}\kappa_{2}(k^{2}-\omega^{2})^{4} & =0,\label{eq:DispScalars}
\end{align}
where 
\begin{equation}
\kappa_{1}:=2c_{1}+c_{2}+c_{4},
\end{equation}
and
\begin{equation}
\kappa_{2}:=4c_{1}^{2}+12c_{3}(c_{2}+c_{4})-3c_{5}^{2}+4c_{1}(c_{2}+4c_{3}+c_{4}+c_{5}).
\end{equation}

The scalar perturbation equations and the dispersion relation are in agreement
with the results of Ref.\cite{BeltranJimenez:2018vdo}. The physical consequence of Eq.(\ref{eq:DispScalars}) is that all the scalar modes propagate with the speed of light, i.e. on the null cone. Also, note that $k^{\mu}k_{\mu}=k^{2}-\omega^{2}$
is the norm of the wave vector related to the scalar perturbation
$h_{\mu\nu}^{S}$ of the metric which emerges by Fourier transforming through $h_{\mu\nu}^{S}(x,t)\rightarrow h_{\mu\nu}^{S}e^{ik_{\mu}x^{\mu}}$. Therefore, if the norm of $k_{\mu}$ is null, then this just means that the
spacetime wave vector $k^{\mu}$ is also null i.e it lies on the
null cone and all of the above relations are equivalent to saying that the scalar
modes of the gravitational wave propagate at the speed of light.

In a similar fashion, we find the field equations describing vector-tensor
perturbations by inserting $h_{\mu\nu}\rightarrow(h_{\mu\nu}^{V}+h_{\mu\nu}^{T})$
into Eq.(\ref{eq:=00003D00003D00003D0003B4E QM semifinal})

\begin{align}
\delta T^{0j} =& (2c_{1}+c_{2}+c_{4})\ddot{B}^{j}+2c_{1}k^{2}B^{j}\label{eq:=0003B4T0jvectensor}\\
 & +(c_{2}+c_{4})k^{2}\dot{E}^{j},\nonumber 
\end{align}

\begin{align}
\delta T^{ij} =& -4c_{1}(\ddot{E}^{ij}+k^{2}E^{ij})\label{eq:=0003B4Tijvectensor}\\
 & -2i\bigl((c_{2}+c_{4})\dot{B}^{(i}k^{j)}\nonumber \\
 & -4ic_{1}\ddot{E}^{(i}k^{j)}-2i(2c_{1}+c_{2}+c_{4})k^{2}E^{(i}k^{j)}.\nonumber 
\end{align}

We now let the energy-momentum tensor vanish since out interest is in determining the dispersion relation for the GW modes in vacuum. Following the same method described for the scalar perturbations, we found
the dispersion relations for the vector modes to be
\begin{equation}
c_{1}\kappa_{1}(k^{2}-\omega^{2})=0,\label{eq:DispVectors}
\end{equation}
and for the tensor modes
\begin{equation}
c_{1}(k^{2}-\omega^{2})=0,\label{eq:DispTensors}
\end{equation}
where again both of these modes propagate in the speed of light. 

Notice that in every dispersion relation the coefficient $c_{1}$
appears explicitly as a multiplying factor which is a result of $c_{1}$
being coupled to the wave operator in Eq.(\ref{eq:=00003D00003D00003D0003B4E QM semifinal}).
Hence, all of the modes propagate with the speed of light, which means that the $E(2)$ framework can be employed to further classify the polarization modes \cite{Eardley:1973br,Eardley:1974nw}. Our
analysis is also in agreement with Ref.\cite{Hohmann:2018wxu} where
another version of the SVT decomposition was employed, and the
$E(2)$ framework used to perform the classification of the polarizations modes for this class of theories. The full polarization modes are classified in that work.

\subsection{GWs in \texorpdfstring{$f(Q)$}{f(Q)} gravity \label{subsec:f(Q)-gravity}}

The other natural generalization of the STEGR scalar is to raise the Lagrangian to an arbitrary function. This is analogous to the $f(\mathcal{R})$ gravity paradigm. Saying that, this is distinct in that the resulting field equations remain second order, meaning that $f(Q)\neq f(\mathcal{R})$. This happens due to the contribution of the boundary term in Eq.(\ref{ricci_nonme_equiv}) which renders $f(Q)$ gravity as a genuinely distinct theory.

Using Eq.(\ref{STG_field_eqn}) we can write the field equations as
\begin{align}
f'(Q) & 
\left[\frac{2\left(\nabla_{\alpha}\sqrt{-g}P^{\alpha}{}_{\mu\nu}\right)}{\sqrt{-g}}+P_{
\mu\alpha\beta}Q_{\nu}{}^{\alpha\beta}-2Q_{\alpha\beta\mu}P^{\alpha\beta}{}_{\nu}\right]
\nonumber \\
 &
-\frac{1}{2}g_{\mu\nu}f(Q)+2P^{\alpha}{}_{\mu\nu}\partial_{\alpha}(f'(Q))=T_{\mu\nu},
\label{eq:f(Q) eq}
\end{align}
where $f'(Q):= df/dQ$. We can rewrite \eqref{eq:f(Q) eq} to
be represented by the Einstein tensor $\mathcal{G}_{\mu\nu}$ as determined
using the Levi-Civita connection, obtaining
\begin{align}
\frac{1}{2} & 
g_{\mu\nu}\left[-f(Q)+f'(Q)Q\right]+f'(Q)\mathcal{G}_{\mu\nu}+2P^{\alpha}{}_{\mu\nu}
\nabla_{\alpha}f'(Q)\nonumber \\
 & \!\!\!\!\!=T_{\mu\nu}.\label{eq:f(Q) eq Einstein}
\end{align}
Following the same procedure as in the irreducible decomposition,
we perturb the metric up to first order using (\ref{metric_pert})
and determine the field equations in Eq.\eqref{eq:f(Q) eq Einstein}
which result in
\begin{align}
\eta_{\mu\nu}f(0) & =0,\label{eq:f(0)}\\
\mathcal{G}_{\mu\nu}^{(1)}f'(0)-\frac{1}{2}h_{\mu\nu}f(0) & =0.\label{eq:wave eq 1}
\end{align}
These are the zeroth and first order perturbation equations which
yield a vanishing cosmological constant, i.e $f(0)=0$ and 
\begin{equation}
\mathcal{G}_{\mu\nu}^{(1)}f'(0)=0,\label{eq:wave eq 2}
\end{equation}
respectively.
These equations are completely equivalent to GR in the first order perturbation
regime, for the non-trivial case $f'(0)\neq0$. This again implies that we acquire
the same speed and polarizations of waves  as in GR.
  Note that if one starts from
$f(\mathbb{Q})$, i.e using the modified version of the generalized
non-metricity scalar of \eqref{eq:Gen Q scalar}, the same result
as in \eqref{eq:wave eq 2} is obtained when the appropriate choices for STEGR are chosen. 

The behaviour of Eqs.(\ref{eq:f(0)},\ref{eq:wave eq 1}) is identical to the 
case of $f(T)$ gravity \cite{Farrugia:2018gyz,Cai:2018rzd}.
The significance of this result is that the general class of $f(Q)$
theories passes the polarization constraints of the
LIGO-Virgo observation of a binary black hole coalescence \cite{Abbott:2017oio}.

\section{Gravitational waves in theories with Scalar field Coupling to \texorpdfstring{$f(Q)$}{f(Q)} Gravity}
\label{secIV}

In this section, we investigate the GWs which arise in the extended theory where a scalar
field $\phi$ is nonminimally coupled to the nonmetricity scalar $Q$, together
with the presence of a coupled kinetic energy and potential. The study
of GWs in the context of scalar-tensor theories has
been investigated in various works, for instance in the  nonminimal coupling
to torsion scalar and boundary term \cite{Abedi:2017jqx}, in scalar-tensor
equivalent of $f(\mathcal{R})$ gravity \cite{Liang:2017ahj,Rizwana:2016qdq}, in
Horndeski theory \cite{Gong:2017bru} and in GR couplings \cite{Wagoner:1970vr} as well as Ho\v{r}ava gravity \cite{Gong:2018vbo} and generalized TeVeS theory \cite{Gong:2018cgj,Horndeski:1974wa}.
In most works, a linearized gravity approach is considered to examine
the properties of the GWs arising from the theory.
Hence, the approach considers metric perturbations around a
Minkowski background as 
\begin{equation}
g_{\mu\nu}=\eta_{\mu\nu}+h_{\mu\nu}+\dots,\label{metric_pert2}
\end{equation}
where $|h_{\mu\nu}|\ll1$, which represents the first-order correction
to the metric as in Eq.(\ref{metric_pert}). For the scalar field, a perturbative approximation
is considered which takes the form 
\begin{equation}
\phi=\phi_{0}(x_{\mu})+\delta\phi(x_{\mu})+\dots,
\end{equation}
where $|\delta\phi(x_{\mu})|\ll1$ and likewise it represents a first-order perturbation. 
We mention that in this work we allow the background scalar field  to be 
space and time dependent
and not necessarily constant. This will allow for a broader analysis
of the resulting perturbed equations of motion comparing to the literature.

The gravitational Lagrangian that we study in this section is the one
considered in \cite{Jarv:2018bgs} 
\begin{equation}
\mathcal{L}_{g}=\mathcal{A}(\phi)Q-\mathcal{B}(\phi)g^{\alpha\beta}\partial_{\alpha}
\phi\partial_{\beta}\phi-2\mathcal{V}(\phi),\label{scal_fiel_coup}
\end{equation}
where $\mathcal{A}$, $\mathcal{B}$ represent the coupling strengths
to the nonmetricity scalar and kinetic energy of the scalar field
respectively, and $\mathcal{V}(\phi)$ is the potential energy of the
scalar field. In the absence of matter fields  the gravitational field equations and the
scalar-field equation are respectively found to be
\begin{align}
0 & 
=\mathcal{A}\mathcal{G}_{\mu\nu}+2\udt{P}{\alpha}{\mu\nu}\partial_{\alpha}\mathcal{A}
+\frac{1}{
2}g_{\mu\nu}\left(\mathcal{B}g^{\alpha\beta}\partial_{\alpha}
\phi\partial_\beta\phi+2\mathcal{V}\right)\nonumber \\
 & \ \ \ -\mathcal{B}\partial_{\mu}\phi\partial_{\nu}\phi,\label{eq:scalar-fieldeq}\\
0 & 
=2\mathcal{B}\mathcal{D}_{\alpha}\mathcal{D}^{\alpha}\phi+\mathcal{B}_{\phi}g^{\alpha\beta
}\partial_{\alpha}\phi\partial_{\beta}\phi+\mathcal{A}_{\phi}Q-2\mathcal{V}_{\phi},
\label{eq:scalar-kgeq}
\end{align}
where a subscript wrt the scalar field represents a derivative, i.e. $\mathcal{B}_{\phi}=d\mathcal{B}/d\phi$.

The next step is to consider perturbations over the equations and solve them order by order. We start with the zeroth order perturbation of Eq.\eqref{eq:scalar-fieldeq}, which yields
\begin{equation}
0=\frac{1}{2}\eta_{\mu\nu}\left(\mathcal{B}^{(0)}\eta^{\alpha\beta}\partial_{\alpha}\phi_{
0}\partial_{\beta}\phi_{0}+2\mathcal{V}^{(0)}\right)-\mathcal{B}^{(0)}\partial_{\mu}\phi_{
0} \partial_{\nu}\phi_{0},
\end{equation}
where superscript bracketed numerals again refer to the order of the perturbation of the quantity. This leads to a system of 10 equations that yield a set of constraints on the potential Lagrangians. The first is that $\mathcal{V}^{(0)}=0$. Then, one of the following scenarios must hold:
\begin{enumerate}
\item $\mathcal{B}^{(0)}=0$; 
\item $\mathcal{B}^{(0)}=0$ and $\phi_{0}=\phi_{0}(x_{\mu})$ for some
$\mu$; 
\item $\phi_{0}=const$. 
\end{enumerate}
Using these conditions, the first-order equation in Eq.\eqref{eq:scalar-fieldeq} simplifies to 
\begin{align}
 & 
0=2\partial_{\alpha}\left(\mathcal{A}^{(0)}\udt{P}{\alpha(1)}{\mu\nu}\right)+\frac{1}{2}
\eta_{\mu\nu}\left[\mathcal{B}^{(1)}\,\eta^{\alpha\beta}\partial_{\alpha}\phi_{0}\partial_{\beta} \phi_{0}+2\mathcal{V}^{(1)}\right]\nonumber \\
 & \ \ \ \
-\mathcal{B}^{(1)}\,\partial_{\mu}\phi_{0}\partial_{\nu}\phi_{0}.
\label{eq:scalar-fieldeq-1st}
\end{align}
On the other hand, the zeroth and first order of the scalar field
equation in Eq.\eqref{eq:scalar-kgeq} result into the following linearized equations
\begin{align}
 & \!\!\!
0=\mathcal{B}_{\phi}^{(0)}\eta^{\mu\nu}\partial_{\mu}\phi_{0}\partial_{\nu}\phi_{0}
-2\mathcal{V}_{\phi}^{(0)},\label{eq:scalar-kgeq-0th}\\
 & \!\!\!
0=2\mathcal{B}^{(0)}\Box\delta\phi+2\mathcal{B}^{(1)}\,\Box\phi_{0}+2\mathcal{B}_{\phi}^{(0)}\,\eta^{
\mu\nu}\partial_{\mu}\phi_{0}\partial_{\nu}\delta\phi\nonumber \\
 & \ \,
-\mathcal{B}_{\phi}^{(0)}h^{\mu\nu}\partial_{\mu}\phi_{0}\partial_{\nu}\phi_{0}+\mathcal{B}^{(1)}_{\phi}\,\eta^{\mu\nu}\partial_{\mu}\phi_{0}\partial_{\nu}\phi_{0}-2\mathcal{V}_{\phi}^{ (1)},\label{eq:scalar-kgeq-1st}
\end{align}
where $\Box:=\eta^{\mu\nu}\partial_{\mu}\partial_{\nu}$ is d'Alembert's operator. We now investigate the three cases separately.

\subsection{\texorpdfstring{$\mathcal{B}^{(0)}=0$}{B(0)=0}}

\noindent For the first case, Eqs.(\ref{eq:scalar-fieldeq-1st},\ref{eq:scalar-kgeq-0th},\ref{eq:scalar-kgeq-1st}) become
\begin{align}
 & 
0=2\partial_{\alpha}\left(\mathcal{A}^{(0)}\udt{P}{\alpha(1)}{\mu\nu}\right)+\frac{1}{2}
\eta_{\mu\nu}\left[\mathcal{B}^{(1)}\,\eta^{\alpha\beta}\partial_{\alpha}\phi_{0}\partial_{
\beta} \phi_{0}+2\mathcal{V}^{(1)}\right]\nonumber \\
 & \ \ \ \
-\mathcal{B}^{(1)}\,\partial_{\mu}\phi_{0}\partial_{\nu}\phi_{0},
\label{eq:scalar-fieldeq-1st-caseI}\\
 & 
0=\mathcal{B}_{\phi}^{(0)}\eta^{\mu\nu}\partial_{\mu}\phi_{0}\partial_{\nu}\phi_{0}
-2\mathcal{V}_
{\phi}^{(0)},\label{eq:scalar-kgdeq-0th-caseI}\\
 & 
0=2\mathcal{B}^{(1)}\,\Box\phi_{0}+2\mathcal{B}_{\phi}^{(0)}\eta^{\mu\nu}\partial_{\mu}\phi_{0}\partial_{\nu}\delta\phi-\mathcal{B}_{\phi}^{(0)}h^{\mu\nu}\partial_{\mu}\phi_{0}\partial_{ \nu}\phi_{0}\nonumber \\
 & \ \ \ \
+\mathcal{B}^{(1)}_{\phi}\,\eta^{\mu\nu}\partial_{\mu}\phi_{0}\partial_{\nu}\phi_{0}-2\mathcal{V}_{\phi}^{(1)}.\label{eq:scalar-kgdeq-1st-caseI}
\end{align}
For this case, not much can be done with the field equations, in order to examine
the behaviour of the perturbations due to the complexity of the system.
Nevertheless, results can be obtained within certain considerations.

If we assume that $\mathcal{B}$ is Taylor expandable around some value $\phi=\phi_\star$, then $\mathcal{B}(\phi)=\sum\limits_{n=0}^{\infty}\frac{\mathcal{B}^{(n)}(\phi_\star)}{n!}\left(\phi-\phi_\star\right)^{n}$. Since $\mathcal{B}^{(0)}=0$, this leaves two possibilities, \textbf{(i)} $\mathcal{B}^{(n)}(\phi_\star)=0$ for every $n$, which implies $\mathcal{B}(\phi) = 0$ (in other words, absent from the Lagrangian), or \textbf{(ii)} $\phi_{0}$ achieves a constant (real or complex) value in terms of the Taylor coefficients and $\phi_\star$. The latter case is not of interest here since the constant case is investigated separately, and hence we investigate only the former case.

The scalar zeroth and first order equations in Eq.\eqref{eq:scalar-kgdeq-0th-caseI}
and Eq.\eqref{eq:scalar-kgdeq-1st-caseI} result in the conditions
$\mathcal{V}_{\phi}^{(0)}=\mathcal{V}_{\phi\phi}^{(0)}=0$. On the
other hand, the first-order field equation in Eq.\eqref{eq:scalar-fieldeq-1st-caseI}
reduces to 
\begin{equation}
0=\partial_{\alpha}\left(\mathcal{A}^{(0)}\udt{P}{\alpha(1)}{\mu\nu}\right).
\label{eq:B0-GW-sol}
\end{equation}
Solving the partial differential equation  for the perturbations $h_{\mu\nu}$ in general
is not possible here unless prior knowledge of $\mathcal{A}^{(0)}$ and
$\phi_{0}$ are known. However, one can make note of the following.
The coupling term $\mathcal{A}$ represents the coupling strength
to STEGR. In most cases we are interested in coupling strengths which
deviate slightly from STEGR, and hence one can assume the form of
$\mathcal{A}$ to be 
\begin{equation}
\mathcal{A}(\phi)=1+\epsilon_2\bar{\mathcal{A}}(\phi)+\mathcal{O}\left(\epsilon_2^{2}\right),
\end{equation}
where $\epsilon_2$ is some small parameter (i.e. $|\epsilon_2|\ll1$)
and $\bar{\mathcal{A}}$ is a function of $\phi$. In this way,
the zeroth-order perturbation takes the form 
\begin{equation}
\mathcal{A}^{(0)}=1+\epsilon_2\bar{\mathcal{A}}^{(0)}+\mathcal{O}\left(\epsilon_2^{2}\right),
\end{equation}
where $\bar{\mathcal{A}}^{(0)}=\bar{\mathcal{A}}(\phi_{0})$.
Thus, instead of solving Eq.\eqref{eq:B0-GW-sol} in general, the
equation is solved perturbatively  order by order in terms
of $\epsilon_2$. This can be achieved by taking a perturbative solution
for $h_{\mu\nu}$ in the form 
\begin{equation}
h_{\mu\nu}=h_{\mu\nu}^{\text{STEGR}}+\epsilon_2 \bar{h}_{\mu\nu}+\mathcal{O}(\epsilon_2^{2}).
\end{equation}
Here, the STEGR superscript denotes the absence of the $\mathcal{A}$ coupling while the bar denotes the first order contribution of the latter, which notation will be assumed in the remainder of this section. Since $\udt{P}{\alpha(1)}{\mu\nu}$
is explicitly dependent on $h_{\mu\nu}$, this leads to a similar
order expansion of the form 
\begin{equation}
\udt{P}{\alpha(1)}{\mu\nu}=P^{\alpha(1)\text{STEGR}}_{\hspace{5mm}\mu\nu}+\epsilon_2\udt{\bar{P}}{\alpha(1)}{\mu\nu}+\mathcal{O}(\epsilon_2^{2}),
\end{equation}
where it can be shown that 
$P^{\alpha(1)\text{STEGR}}_{\hspace{5mm}\mu\nu}=\udt{P}{\alpha(1)}{\mu\nu}|_{h_{\mu\nu}\rightarrow 
h_{\mu\nu}^{\text{STEGR}}}$
and $\udt{\bar{P}}{\alpha(1)}{\mu\nu}=\udt{P}{\alpha(1)}{\mu\nu}|_{h_{\mu\nu}\rightarrow 
\bar{h}_{\mu\nu}}
$.
Therefore, expanding Eq.\eqref{eq:B0-GW-sol} order by order yields
the iterative system of equations
\begin{align}
0 & =\partial_{\alpha}P^{\alpha(1)\text{STEGR}}_{\hspace{5mm}\mu\nu},\label{eq:B0-GW-sol-0th}\\
0 & 
=\partial_{\alpha}\left(\udt{\bar{P}}{\alpha(1)}{\mu\nu}+\bar{\mathcal{A}}^{(0)}P^{\alpha(1)\text{STEGR}}_{\hspace{5mm}\mu\nu}\right) \nonumber \\
 & 
=\partial_{\alpha}\udt{\bar{P}}{\alpha(1)}{\mu\nu}+P^{\alpha(1)\text{STEGR}}_{\hspace{5mm}\mu\nu} \partial_{\alpha}
\bar{\mathcal{A}}^{(0)},\label{eq:B0-GW-sol-1st}
\end{align}
where in Eq.\eqref{eq:B0-GW-sol-1st} we used Eq.\eqref{eq:B0-GW-sol-0th}. One can easily observe that the solution for Eq.\eqref{eq:B0-GW-sol-0th}
yields the standard STEGR GW solution, while from Eq.\eqref{eq:B0-GW-sol-1st}
the first-order correction $\bar{h}_{\mu\nu}$ depends on the STEGR
solution and the scalar field coupling which together act as a source
term. Hence, the choice of the coupling strength is important as it affects the corrections to the standard STEGR GW modes. Furthermore, it is necessary for the scalar field to be strictly non-constant,
otherwise the modes reduce to those of STEGR (since $\bar{\mathcal{A}}^{(0)}$
would become constant and hence the source term would become zero)
as expected.

\subsection{\texorpdfstring{$\mathcal{B}^{(0)}=0$}{B(0)=0} and \texorpdfstring{$\phi_{0}=\phi_{0}(x_{\mu})$}{phi} (for some \texorpdfstring{$\mu$)}{mu}}

For these conditions, Eqs.(\ref{eq:scalar-fieldeq-1st},\ref{eq:scalar-kgeq-0th},\ref{eq:scalar-kgeq-1st}) reduce to 
\begin{align}
 & 
0=2\partial_{\alpha}\left(\mathcal{A}^{(0)}\udt{P}{\alpha(1)}{\mu\nu}\right)+\frac{1}{2}
\eta_{\mu\nu}\left[\mathcal{B}^{(1)}\,\eta^{\rho\rho}\left(\partial_{\rho}\phi_{0}\right)^{2
} +2\mathcal{V}^{(1)}\right]\nonumber \\
 & \ \ \ \
-\mathcal{B}^{(1)}\,\partial_{\mu}\phi_{0}\partial_{\nu}\phi_{0},
\label{eq:scalar-fieldeq-1st-caseII}\\
 & 
0=\mathcal{B}_{\phi}^{(0)}\eta^{\rho\rho}\left(\partial_{\rho}\phi_{0}\right)^{2}
-2\mathcal{V}_{\phi}^{(0)},\label{eq:scalar-kgdeq-0th-caseII}\\
 & 
0=2\mathcal{B}^{(1)}\,\Box\phi_{0}+2\mathcal{B}_{\phi}^{(0)}\eta^{\rho\rho}\partial_{\rho}
\phi_{0}\partial_{\rho}\delta\phi-\mathcal{B}_{\phi}^{(0)}h^{\rho\rho}\left(\partial_{\rho
}\phi_{0} \right)^{2}\nonumber \\
 & \ \ \ \
+\mathcal{B}^{(1)}_{\phi}\,\eta^{\rho\rho}\left(\partial_{\rho}\phi_{0}\right)^{2}-2\mathcal
{V}_{\phi}^{(1)},\label{eq:scalar-kgdeq-1st-caseII}
\end{align}
where $\Box\phi_{0}=\eta^{\rho\rho}\partial_{\rho}^{2}\phi_{0}$.
In what follows, we assume that $\phi_{0}$ is non-constant, since in that case
the model reduces to the next case $\phi_{0}=const$, analyzed in the next subsection. 

Taking the derivative  of Eq.\eqref{eq:scalar-kgdeq-0th-caseII} with respect to 
$x_{\rho}$, and using the fact that $\mathcal{C}^{(1)}=\mathcal{C}_{\phi}^{(0)}\delta\phi$
for any $\mathcal{C}(\phi)$, Eq.\eqref{eq:scalar-kgdeq-1st-caseII} reduces
to 
\begin{equation}
0=\mathcal{B}_{\phi}^{(0)}\left[2\eta^{\rho\rho}\partial_{\rho}\delta\phi-h^{\rho\rho}
\partial_{\rho}\phi_{0}\right].
\end{equation}
This leaves two possibilities, either $\mathcal{B}_{\phi}^{(0)}=0$
or $2\eta^{\rho\rho}\partial_{\rho}\delta\phi-h^{\rho\rho}\partial_{\rho}\phi_{0}=0$.
In the former case, Eq.\eqref{eq:scalar-fieldeq-1st-caseII} reduces to 
\begin{equation}
0=\partial_{\alpha}\left(\mathcal{A}^{(0)}\udt{P}{\alpha(1)}{\mu\nu}\right).
\end{equation}
As discussed in the previous subsection, the solution for the metric
perturbation cannot be obtained in general but its behaviour can be
examined provided that the coupling strength $\mathcal{A}$ can be
expanded as STEGR with a small correction.

In the remaining case, provided that the behaviour for $\phi_{0}$
is known, this leads to the system
\begin{align}
 & 
0=2\partial_{\alpha}\left(\mathcal{A}^{(0)}\udt{P}{\alpha(1)}{\mu\nu}\right)+\left[2\eta_{
\mu\nu}
\mathcal{V}_{\phi}^{(0)}-\mathcal{B}_{\phi}^{(0)}\partial_{\mu}\phi_{0}\partial_{\nu}\phi_
{0}\right]
\delta\phi,\\
 & 0=2\eta^{\rho\rho}\partial_{\rho}\delta\phi-h^{\rho\rho}\partial_{\rho}\phi_{0},
\end{align}
where the field equation in question was simplified using Eq.\eqref{eq:scalar-kgdeq-0th-caseII}.
However, this case cannot be analytically solved for both $\delta\phi$
and $h_{\mu\nu}$, and hence will not be investigated further. Nonetheless,
we mention that if we assume that $\mathcal{B}$ is Taylor expandable
around $\phi=0$ as in the previous case, then this instance would not appear
and the previous case would follow.

\subsection{\texorpdfstring{$\phi_{0}=const.$}{phi}}

For the constant case, Eqs.(\ref{eq:scalar-fieldeq-1st},\ref{eq:scalar-kgeq-0th},\ref{eq:scalar-kgeq-1st}) simplify to 
\begin{align}
 & 
0=2\partial_{\alpha}\left(\mathcal{A}^{(0)}\udt{P}{\alpha(1)}{\mu\nu}\right)+\eta_{\mu\nu}
\mathcal{V}^{(1)},\label{eq:scalar-fieldeq-1st-caseIII}\\
 & 0=\mathcal{V}_{\phi}^{(0)},\label{eq:scalar-kgeq-0th-caseIII}\\
 & 
0=\mathcal{B}^{(0)}\Box\delta\phi-\mathcal{V}_{\phi}^{(1)}.
\label{eq:scalar-kgdeq-1st-caseIII}
\end{align}
As $\mathcal{V}^{(1)}=\mathcal{V}_{\phi}^{(0)}\delta\phi$, 
Eq.\eqref{eq:scalar-fieldeq-1st-caseIII}
simplifies further to 
\begin{equation}
0=\partial_{\alpha}\left(\mathcal{A}^{(0)}\udt{P}{\alpha(1)}{\mu\nu}\right).
\end{equation}
Furthermore, since $\phi_{0}$ is constant, every zeroth-order quantity
of functions of $\phi$ will be constant. Therefore, since $\mathcal{A}^{(0)}$
is constant, this simplifies the expression to 
$0=\mathcal{A}^{(0)}\mathcal{G}_{\mu\nu}^{(1)}$.
Since the $\mathcal{A}$ coupling has to be non-zero (otherwise no
STEGR contributions appear), this reduces to the standard STEGR perturbation
equation. On the other hand, since 
$\mathcal{V}_{\phi}^{(1)}=\mathcal{V}_{\phi\phi}^{(0)}\delta\phi$,
Eq.\eqref{eq:scalar-kgdeq-1st-caseIII} becomes 
\begin{equation}
\mathcal{B}^{(0)}\Box\delta\phi-\mathcal{V}_{\phi\phi}^{(0)}\delta\phi=0,
\end{equation}
which yields a wave equation solution with effective mass 
\begin{equation}
m^{2}=\frac{\mathcal{V}_{\phi\phi}^{(0)}}{\mathcal{B}^{(0)}}.
\end{equation}
Therefore, the scalar field evolves independently of the metric perturbations.

In summary, we can deduce that while the GW analysis of this section contains three subcases, the main distinction comes from when the background value of the scalar field is a constant. If it is, then the metric propagates with two polarizations identical to GR with an independent massive mode in the scalar field. This means that we recover STEGR, while in the scenario where the background value of the scalar field is not constant, the scalar field acts as a source term that attenuates the GW signature but does not produce any extra polarization modes.

\section{Gravitational waves in \texorpdfstring{$f(Q,B)$}{f(Q,B)} Gravity}
\label{secV}

In this section we investigate GWs within the $f(Q,B)$ gravity context, which is a more 
general scenario than $f(Q)$ with a Lagrangian that depends
both on the nonmetricity scalar and the boundary term that forms with the Levi-Civita Ricci scalar as shown in Eq.(\ref{ricci_nonme_equiv}). In essence, this is a generalization of $f(\mathcal{R})$ gravity in terms of the order contributions to the Lagrangian.

\subsection{\texorpdfstring{$f(Q,B)$}{f(Q,B)} gravity}

\noindent The action of  $f(Q,B)$ gravity naturally writes as 
\begin{align}
S&=\frac{1}{16\pi G}\int d^{4}x \big[\sqrt{-g}f(Q,B) + \dut{\lambda}{\alpha}{\beta\mu\nu}\udt{R}{\alpha}{\beta\mu\nu}\nonumber\\
&+\dut{\lambda}{\alpha}{\mu\nu}\udt{T}{\alpha}{\mu\nu}\big] + \int d^{4}x\sqrt{-g}\mathcal{L}_{m},
\end{align}
where $\mathcal{L}_{m}$ refers to any source contributions, and the Lagrange multipliers are eliminated in the same way as in Eq.(\ref{connection_field_eq}) which is again solved by adopting the coincident gauge. This is interesting since besides being another potential generalization
of STEGR, it also offers an attractive alternative interpretation
of the well studied $f(\mathcal{R})$ modification of GR. This implies
that despite the plethora of work on the topic, $f(Q,B)$ gravity
offers an alternative direction for a broader class of $f(\mathcal{R})$
equivalent theories of gravity.

Variation of the action with respect to  the metric yields the field equations
\begin{align}
 & 
\frac{1}{2}g_{\mu\nu}\left[-f(Q,B)+f_{B}B\right]+2P_{~\mu\nu}^{\alpha}\partial_{\alpha}(f_
{Q}+f_{B})\nonumber \\
 & 
+f_{Q}\left[\frac{2}{\sqrt{-g}}\nabla_{\alpha}\left(\sqrt{-g}P_{~\mu\nu}^{\alpha}
\right)+P_{\mu\alpha\beta}Q_{\nu}^{~\alpha\beta}-2Q_{\alpha\beta\mu}P_{~~~\nu}^{
\alpha\beta}\right] \nonumber \\
 & 
+\mathcal{D}_{\mu}\mathcal{D}_{\nu}f_{B}-g_{\mu\nu}\mathcal{D}^{\alpha}\mathcal{D}_{\alpha
}f_{B}
=8\pi GT_{\mu\nu},
\label{f_q_b_field_eqns}
\end{align}
where $\nabla_{\mu}$ is the STG covariant derivative, $\mathcal{D}_{\mu}$
is the Levi-Civita covariant derivative,  and $P_{~\mu\nu}^{\alpha}$ is the 
superpotential defined
in  (\ref{conj_Lag_ten}), which now becomes
\begin{align}
P_{~\mu\nu}^{\alpha}= & 
\frac{1}{4}\Big[-Q^{\alpha}{}_{\mu\nu}+2Q_{(\mu\phantom{\alpha}\nu)}^{\phantom{\mu}\alpha}
+Q^{\alpha}g_{\mu\nu}\nonumber \\
 & ~~~~-\tilde{Q}^{\alpha}g_{\mu\nu}-\delta_{(\mu}^{\alpha}Q_{\nu)}\Big].
\end{align}
Using the GR limit $f(Q,B)\rightarrow\mathcal{R}=Q-B$, we can identify
the Einstein tensor in the field equations and write 
\begin{align}
 & \frac{1}{2}g_{\mu\nu}\left[-f(Q,B)+f_{Q}Q+f_{B}B\right]\nonumber \\
 & +f_{Q}\mathcal{G}_{\mu\nu}+2P_{~\mu\nu}^{\alpha}\partial_{\alpha}(f_{Q}+f_{B})\nonumber 
\\
 & 
+\left[\mathcal{D}_{\mu}\mathcal{D}_{\nu}-g_{\mu\nu}\mathcal{D}^{\alpha}\mathcal{D}_{
\alpha}\right]f_{B}=8\pi GT_{\mu\nu},\label{Feq}
\end{align}
where $\mathcal{G}_{\mu\nu}=\mathcal{R}_{\mu\nu}-\frac{1}{2}g_{\mu\nu}\mathcal{R}$
is the Einstein tensor calculated using Levi-Civita connection.

We can retrieve the field equation of $f(\mathcal{R})$ gravity \cite{DeFelice:2010aj}
by taking the limit $f(Q,B)\rightarrow f(\mathcal{R}=Q-B)$, where
we have $f_{B}\rightarrow-f_{\mathcal{R}}$, $f_{Q}\rightarrow f_{\mathcal{R}}$
and $f_{BB}\rightarrow f_{\mathcal{R}\mathcal{R}}$, yielding 
\begin{align}
 & -\frac{1}{2}g_{\mu\nu}f(\mathcal{R})+f_{\mathcal{R}}\mathcal{R}_{\mu\nu}\nonumber \\
 & 
+\left[g_{\mu\nu}\mathcal{D}^{\alpha}\mathcal{D}_{\alpha}-\mathcal{D}_{\mu}\mathcal{D}_{
\nu}\right]f_{\mathcal{R}}=8\pi GT_{\mu\nu},
\end{align}
which agree with the $f(\mathcal{R})$ gravity field equations as expected.

\subsection{GWs in \texorpdfstring{$f(Q,B)$}{f(Q,B)} gravity}

We now proceed to the study of GWs within the context of $f(Q,B)$ theory. As before, we assume the coincident
gauge where the connection vanishes \cite{BeltranJimenez:2017tkd}.
As a further coincidence, the field equations that emerge in Eq.(\ref{f_q_b_field_eqns})
are identical to those in the teleparallel case under the symbolic
change $T\rightarrow Q$ and superpotential change, which was studied
in Ref.\cite{Farrugia:2018gyz}. Furthermore, the $Q$ and $T$ scalars
are both second-order quantities, thus at first order only the boundary
terms of the theories will contribute. The foundations of these theories
are wholly distinct from each other but given their relation to the
Ricci scalar the above result is not completely unexpected.

Consider the metric perturbation of Eq.\eqref{metric_pert}. At first order, only the field equations that contain the boundary term will survive beyond the cosmological constant. This can be related to the first order of the Ricci scalar using Eq.(\ref{ricci_nonme_equiv}), which gives 
\begin{align}
B^{(1)}= & 
\left(\mathcal{D}_{\alpha}(Q^{\alpha}-\tilde{Q}^{\alpha})\right)^{(1)},\nonumber \\
= & 
\partial_{\alpha}(\eta^{\beta\gamma}\partial^{\alpha}h_{\beta\gamma}-\eta^{\alpha\beta}
\partial^{\gamma}h_{\beta\gamma}),\nonumber \\
= & \square h_{~\beta}^{\beta}-\partial^{\alpha}\partial^{\beta}h_{\alpha\beta}=-\mathcal{R}^{(1)}.
\end{align}
We assume a Taylor expansion of $f(Q,B)$ around $(0,0)$, so that the expansion of the Lagrangian takes the form
\begin{align}
f(Q,B)= & f(0,0)+f_{Q}(0,0)Q+f_{B}(0,0)B+\frac{1}{2}f_{QQ}(0,0)Q^{2}\nonumber \\
 & +\frac{1}{2}f_{BB}(0,0)B^{2}+f_{QB}QB+\cdots\,.
\end{align}
We then consider  a vacuum background where $T_{\mu\nu}=0$ and writing the field equations up to first order as 
\begin{align}
\eta_{\mu\nu}f(0,0)= & 0,\\
f_{Q}(0,0)\mathcal{G}_{\mu\nu}^{(1)}-f_{BB}(0,0)(\partial_{\mu}\partial_{\nu}-\eta_{\mu\nu
}\square)\mathcal{R}^{(1)}= & 0,\label{f_q_b_first_ord_eqn}
\end{align}
where we used the solution of the first equation $f(0,0)=0$ is used to simplify the second equation.

\noindent Taking the trace of Eq.\eqref{f_q_b_first_ord_eqn} yields 
\begin{align}
-f_{Q}(0,0)\mathcal{R}^{(1)}+3f_{BB}(0,0)\square\mathcal{R}^{(1)}=0.\label{R_klein_eq}
\end{align}
We identify a Klein Gordon type equation $(\square-m^{2})\mathcal{R}^{(1)}=0$,
where the effective mass is given 
\begin{equation}
m^{2}=\frac{f_{Q}(0,0)}{3f_{BB}(0,0)}.\label{mass_R}
\end{equation}

From Eq.\eqref{R_klein_eq}, it is evident that a massive wave propagates in the GW signature, but the wave is composed of at most six polarizations and so no further information can be extracted from this scalar wave equation. Given that the Klein-Gordon equation does not operate directly $h_{\mu\nu}$ but on $\mathcal{R}^{(1)}=\square h_{~\beta}^{\beta}-\partial^{\alpha}\partial^{\beta}h_{\alpha\beta}$, the massive mode is actually an expression of the Ricci scalar and not the underlying metric perturbation.

As already discussed in \S.\ref{secIIIA}, we do not have the freedom to further set gauge conditions such as the Lorenz gauge since we have already chosen the coincident gauge for the connection, which permeates even at perturbative level. This means that we must consider again an SVT decomposition of the linearized equations. Structurally, this is very similar to $f(\mathcal{R})$ (see appendix \ref{subsec:f(R)}) so we would expect to find breathing and even longitudinal modes
in $f(Q,B)$, since $f(\mathcal{R})\subseteq f(Q,B)$. In terms of $h_{\mu\nu}$, the linearized field equations for $f(Q,B)$ turn out as
\begin{widetext}
\begin{align}
\tfrac{1}{2}\left(-\square h_{\mu\nu}+g_{\mu\nu}(-\partial_{\lambda}\partial_{\alpha}h^{\alpha\lambda}+\square h)+\partial_{\mu}\partial_{\alpha}h_{\nu}{}^{\alpha}+\partial_{\nu}\partial_{\alpha}h_{\mu}{}^{\alpha}-\partial_{\nu}\partial_{\mu}h\right)\nonumber \\
+\left(\eta_{\mu\nu}(\square\partial_{\beta}\partial_{\alpha}h^{\alpha\beta}-\square^{2}h)-\partial_{\nu}\partial_{\mu}\partial_{\lambda}\partial_{\alpha}h^{\alpha\lambda}+\partial_{\nu}\partial_{\mu}\partial_{\lambda}\partial^{\lambda}h\right)C_{B} & =0,\label{eq:lineq f(Q,B) h}
\end{align}
\end{widetext}
where $f_{Q}:=f_{Q}(0,0)\neq0$, $f_{BB}:=f_{BB}(0,0)\neq0$,
$C_{B}:=f_{BB}/f_{Q}$ and $h=h_{\nu}{}^{\nu}$. It is important to
note that the Lagrangian limits to $f(\mathcal{R})$ and not to $GR$ due to the restrictions on the values of the arbitrary function $f$, which stems from the fact that $f_{BB}$ is present. The choice $f_{Q}\rightarrow f_{\mathcal{R}}$ and $f_{BB}\rightarrow0$  is the limit to $f(\mathcal{R})$
only, since a limit to $GR$ would additionally require that $f_{B}\rightarrow0$.

In STG, the connection is changed from the Levi-Civita to the disformation connection. However, this framework still exists within the broader Riemannian geometry setting and so certain elements can be adopted straightforwardly as in modifications of GR. One such setup is that of a locally freely falling observer on a Riemann manifold making measurements on an incoming GW \cite{WillBook}. This analysis leads directly to the $E(2)$ framework which can consistently be utilized to determine the polarization modes of incoming GWs. This approach is introduced for convenience in appendix \ref{sec:Appendix-Part-Polarizations}. Here, we will distinguish between an extra mode that propagates as a null and almost null wave since the results differ for either case.

In the null wave case, the only non-trivial components from Eq.(\ref{eq:lineq f(Q,B) h})
are
\begin{alignat}{2}
\mathcal{H}_{\mathit{m}\bar{\mathit{m}}} & =-\tfrac{1}{2}\mathit{\ddot{h}}_{\ell\ell} & \Rightarrow\Psi_{2}=0,\\
\mathcal{H}_{\mathit{n}\bar{\mathit{m}}}=\overline{\mathcal{H}_{\mathit{n}\mathit{m}}} & =-\tfrac{1}{2}\mathit{\ddot{h}}_{\ell\bar{\mathit{m}}} & \Rightarrow\Psi_{3}=0,\\
\mathcal{H}_{\mathit{n}\mathit{n}} & =-C_{B}\ddddot{\mathit{h}_{\ell\ell}}-\mathit{\ddot{h}}_{\mathit{m}\bar{\mathit{m}}} & \;\;\;\Rightarrow\Phi_{22}\neq0.
\end{alignat}
which is a result identical to the corresponding case for $f(\mathcal{R})$ theory, and is also consistent with the fact that $f(\mathcal{R})\subseteq f(Q,B)$. 

In the case of an almost null wave, the norm of the wave vector, $l^{\mu}$, gives
\begin{equation}
    \eta_{\mu\nu}\widetilde{\ell}^{\mu}\widetilde{\ell}^{\nu}=\varepsilon\ll 1,
\end{equation}
where $\epsilon$ is related to to change in the propagation speed. We then obtain the following system of equations
\begin{widetext}
\begin{align}
\mathcal{H}_{\mathit{n}\mathit{n}} & =-C_{B}\overset{....}{\mathit{h}_{\ell\ell}}-\mathit{h}_{\mathit{m}\bar{\mathit{m}}}+\varepsilon\bigl(-\tfrac{1}{2}\mathit{h}_{\mathit{m}\bar{\mathit{m}}}+C_{B}(-\overset{....}{\mathit{h}_{\ell\ell}}+3\overset{....}{\mathit{h}_{\ell\mathit{n}}}-2\overset{....}{\mathit{h}_{\mathit{m}\bar{\mathit{m}}}})+\mathit{h}_{\mathit{n}\mathit{n}}\bigr)\\
\mathcal{H}_{\mathit{n}\bar{\mathit{m}}}=\overline{\mathcal{H}_{\mathit{m}\mathit{n}}} & =-\tfrac{1}{2}\mathit{h}_{\ell\bar{\mathit{m}}}+\tfrac{1}{4}\varepsilon(-\mathit{h}_{\ell\bar{\mathit{m}}}+3\mathit{h}_{\mathit{n}\bar{\mathit{m}}})\\
\mathcal{H}_{\mathit{m}\bar{\mathit{m}}} & =-\tfrac{1}{2}\mathit{h}_{\ell\ell}+\varepsilon(-\tfrac{1}{4}\mathit{h}_{\ell\ell}-C_{B}\overset{....}{\mathit{h}_{\ell\ell}}+\tfrac{3}{2}\mathit{h}_{\ell\mathit{n}}-\tfrac{1}{2}\mathit{h}_{\mathit{m}\bar{\mathit{m}}})\\
\mathcal{H}_{\ell\mathit{n}} & =\varepsilon(\tfrac{3}{2}C_{B}\overset{....}{\mathit{h}_{\ell\ell}}-\mathit{h}_{\ell\mathit{n}}+\tfrac{3}{2}\mathit{h}_{\mathit{m}\bar{\mathit{m}}})\\
\mathcal{H}_{\ell\ell} & =\varepsilon\mathit{h}_{\ell\ell}\\
\mathcal{H}_{\ell\bar{\mathit{m}}}=\overline{\mathcal{H}_{\ell\mathit{m}}} & =\tfrac{3}{4}\varepsilon\mathit{h}_{\ell\bar{\mathit{m}}}\\
\mathcal{H}_{\bar{\mathit{m}}\bar{\mathit{m}}}=\overline{\mathcal{H}_{\mathit{m}\mathit{m}}} & =\tfrac{1}{2}\varepsilon\mathit{h}_{\bar{\mathit{m}}\bar{\mathit{m}}}
\end{align}
\end{widetext}
which are again identical to the corresponding one for $f(\mathcal{R})$, when the substituting $C_{B}\leftrightarrow C_{\mathcal{R}}$ is assumed. Therefore, the polarization amplitudes are $\Psi_{2}=0,$ $\Psi_{3}=0$, $\Psi_{4}=0$
but $\Phi_{22}\neq0$. Just as in $f(\mathcal{R})$, by keeping terms
of the form $\mathcal{O}(\varepsilon h_{\mu\nu})$, which are practically
second order, we see that only the breathing scalar mode survives.
One outcome of this analysis is that the propagating degrees
of freedom of $f(Q,B)$ should at least match those of $f(\mathcal{R})$.

The fact that we need to distinguish two cases for the mass of the
total GW has to do with the form of the dispersion
relation of the total wave since both the null and non null cases are potential solutions. It turns out that there is a massive scalar propagating degree of freedom in the class of $f(Q,B)$ theories,
which at first glance, can be observed from Eq.\eqref{R_klein_eq}. Although
the effective mass in Eq.\eqref{mass_R} does not coincide with the (small)
mass imposed by $\eta_{\mu\nu}\widetilde{\ell}^{\mu}\widetilde{\ell}^{\nu}=\varepsilon$
i.e the almost null wave. This merely suggests that the total wave
cannot be studied just by using one dispersion relation but rather
two of them, one for the tensor wave $\eta_{\mu\nu}\widetilde{\ell}^{\mu}\widetilde{\ell}^{\nu}=\varepsilon\equiv0$
and one for the massive scalar wave $\eta_{\mu\nu}\widetilde{\ell}^{\mu}\widetilde{\ell}^{\nu}=\varepsilon<<1$.
In this sense, the theory itself cannot be classified as a whole but
rather component wise. Considering only the tensor modes, $f(Q,B)$ gravity appears as an $N_{3}$ quasi-Lorentz invariant class. On the other hand, considering only the small masssive scalar part $f(Q,B)$ gravity is of $O_{1}$ quasi-Lorentz
invariant class. This hold translates identically for $f(\mathcal{R})$ gravity.

\section{Conclusions}
\label{Conclusions}

In this work we explored the possibility of GWs in STG theories and their extensions, where gravitation is expressed through nonmetricity rather than curvature or torsion of the manifold connection. This form of gravity can be constructed to be equivalent to GR at the level of field equations through  relation in Eq.\eqref{ricci_nonme_equiv}, namely STEGR. However, the boundary term, $B$, renders most generalizations distinct from their GR analogue. Even at the level of STEGR, there are a number of advantages that STG offers, that do not appear in the Levi-Civita connection form of the theory, such as a well-defined energy-momentum tensor for the gravitational field.

We first investigated the GW signature of the general irreducible setting of the STEGR scalar, $\mathbb{Q}$, which represents a novel generalization that does not appear for GR. This is interesting since it may offer a guide to why STEGR (or GR) should at least form part of any modified theory of gravitational Lagrangian. In the general linear case, we derived the linearized field equations in Eq.\eqref{eq:=00003D00003D00003D0003B4E QM semifinal}, and then the dispersion relations for the scalar, vector and tensor modes which all propagate at the speed of light. However, extra polarization modes beyond the two tensor modes of GR exist in certain setting (the full classification is carried out in Ref.\cite{Hohmann:2018wxu}).

Alternatively, modifying the STEGR Lagrangian itself, analogous to the $f(\mathcal{R})$ paradigm results in generally second order field equations. In Eq.\eqref{eq:wave eq 1}, we find that the polarization modes turn out to be identical to those of GR. This means that GW polarization tests cannot distinguish between GR and $f(Q)$ theories of gravity, while the linear irreducible form of STG theory does produce distinct results that would emerge in GW observations.

Furthermore, we investigated the GWs in theories with a nonminimal coupling between $f(Q)$ and a scalar field, where the perturbations were taken at the level of the metric as well as the scalar field itself. While the GW analysis contains three subcases, the conclusions depend on whether background value of the scalar field is constant. If this is constant then two polarization modes of GR propagate in the metric and the scalar field forms a massive mode that decouples from the two polarizations of STEGR. This implies that the massive mode evolves independently of the metric perturbations, and as a consequence would in general propagate with a different speed. In the other case, the background value of the scalar field is not constant, and effectively acts as a source term that attenuates the two polarizations of STEGR but does not produce any extra polarization modes. 

Finally, we analyzed the $f(Q,B)$ gravitational modification which is a further generalization of the analogous $f(\mathcal{R})$ theory, where the nonmetricity scalar $Q$ embodies the second order contribution and the boundary term $B$ the fourth order contributions. Thus, $f(Q,B)$ gravity offers a wider range of Lagrangians to be explored against observations. In our treatment we adopt the $E(2)$ framework to investigate the polarization modes of the propagating GW modes. We find that the two tensor modes appear in exactly the same way as they do in GR, while an extra scalar breathing mode which may appear as a massive mode. In such cases the scalar mode does not propagate at the speed of light.

In general, GW polarizations offer a way of constraining the strong field behavior of any theory of gravity. This is helpful in constructing a realistic theory of gravity. STG and its extensions offer a way for a paradigm shift in our perspective of gravity.

\section*{Acknowledgements}

We thank Jose Beltr\'{a}n Jim\'{e}nez and Manuel Hohmann for useful and insightful conversations on aspects of
this work. The research work disclosed in this paper is partially funded by the
ENDEAVOUR Scholarships Scheme. This article is based upon work from
CANTATA COST (European Cooperation in Science and Technology) action
CA15117, EU Framework Programme Horizon 2020.

\appendix

\section{Scalar Vector Tensor (SVT) Decomposition\label{subsec:Scalar-Vector-Tensor}}

The SVT decomposition of the metric perturbation $h_{\mu\nu}$ assumes the following form
\begin{equation}
h_{\mu\nu}=h_{\mu\nu}^{S}+h_{\mu\nu}^{V}+h_{\mu\nu}^{T},\label{eq:SVT decomp h}
\end{equation}
where
\begin{equation}
h_{\mu\nu}^{S}:=\left(\begin{array}{cc}
-2\varphi & \partial_{i}B\\
\partial_{i}B & -2\psi\delta_{ij}+\partial_{i}\partial_{j}E
\end{array}\right),\label{eq:hS}
\end{equation}
\begin{equation}
h_{\mu\nu}^{V}:=\left(\begin{array}{cc}
0 & B_{i}\\
B_{i} & 2\partial_{(i}E_{j)}
\end{array}\right),\label{eq:hV}
\end{equation}
\begin{equation}
h_{\mu\nu}^{T}:=\left(\begin{array}{cc}
0 & 0\\
0 & 2E_{ij}
\end{array}\right),\label{eq:hT}
\end{equation}
where $E_{ij}$ is symmetric, and $E_{ij}\delta^{ij}=\partial^{i}E_{ij}=0$. Here $\varphi,\psi,B,\,E$ embody the scalar degrees of freedom (DoFs), $B_i,\,E_i$ the vector DoFs, and $E_{ij}$ the tensor DoFs.

In a similar fashion, the perturbation of the energy-momentum tensor can be divided as
\begin{equation}
\delta T^{ij}=\delta p\delta^{ij}+\Sigma^{ij}=\overline{p}\left(\frac{\delta p}{\overline{p}}\delta^{ij}+\Pi^{ij}\right),\label{eq:=0003B4=0003A4ij}
\end{equation}
\begin{equation}
\Sigma^{ij}:=\delta T^{ij}-\frac{1}{3}\delta^{ij}\delta T^{p}{}_{p},\label{eq:=0003A3ij}
\end{equation}
\begin{equation}
\Pi^{ij}:=\frac{\Sigma^{ij}}{\overline{p}}=\frac{1}{\overline{p}}\left(\delta T^{ij}-\frac{1}{3}\delta^{ij}\delta T^{p}{}_{p}\right),\label{eq:=0003A0ij}
\end{equation}
where $\delta p\equiv\delta T^{i}{}_{i}$ is the perturbation of the
pressure $p$, $\overline{p}$ is the mean value of $p$, and $\Sigma^{ij}$
is the traceless and symmetric tensor part of $\delta T^{ij}$, called
the anisotropic pressure. 

In general, $\delta T^{\mu\nu}$ can be constructed from
the scalar perturbations $\delta p,$ $\delta\rho$, the 3-velocity
vector and $\Pi^{ij}$. We can further perform an SVT decomposition
of $\Pi^{ij}$ as 
\begin{eqnarray}
\Pi^{ij} & = & \Pi_{S}^{ij}+\Pi_{V}^{ij}+\Pi_{T}^{ij},\label{eq:=0003A0ij SVT}\\
\Pi_{S}^{ij} & := & \left(\partial^{i}\partial^{j}-\frac{1}{3}\delta^{ij}\nabla^{2}\right)\sigma,\label{eq:=0003A0 scalar}\\
\Pi_{V}^{ij} & := & -\frac{1}{2}\left(\Pi^{i,j}-\Pi^{j,i}\right).\label{eq:=0003A0 vector}\\
\delta_{ip}\Pi_{T}^{ij,p} & = & 0,\label{eq:=0003A0 tensor}
\end{eqnarray}
which is useful for the perturbation decomposition of the individual theories.

\section{\texorpdfstring{$E(2)$}{E(2)} Framework\label{sec:Appendix-Part-Polarizations}}

We will follow the conventions and methods presented in \cite{WillBook}. We assume a local Lorentz frame in which a plane wave traveling in the $Z-$direction is described by the function of retarted time $u\equiv t-Z$
and a wave traveling in the opposite $-Z-$direction is described by
the function of advanced time $v\equiv t+Z$. The functions $u$ and
$v$ define our Newman-Penrose (N-P) basis as follows
\begin{align}
\ell_{\mu} & :=-u_{,\mu}, & n_{\mu}:= & \frac{1}{2}v_{,\mu},
\end{align}
which made local Fermi coordinates by also including the normals $m^{\mu}$
and $\overline{m}^{\mu}$ which comprises the full complex null basis
\begin{align}
\ell^{\mu} & =(1,0,0,0), & n^{\mu} & =\frac{1}{2}(1,0,0,-1),\\
m^{\mu} & =\frac{1}{\sqrt{2}}(0,1,i,0), & \overline{m}^{\mu} & =\frac{1}{\sqrt{2}}(0,1,-i,0).
\end{align}

Assuming a plane wave traveling in the $Z-$direction, the perturbation
of the metric is a function of the retarted time $h_{\mu\nu}=h_{\mu\nu}(u)$.
In this setup, the only possible polarization amplitudes $\Psi_{2},\Psi_{3},\Psi_{4},\Phi_{22}$
can be expressed as\begin{widetext}
\begin{equation}
\begin{array}{lllllcc}
\Psi_{2} & = & -\frac{1}{6}\mathcal{R}_{n\ell n\ell}+O(\varepsilon\mathcal{R}) & = & -\frac{1}{6}\mathcal{R}_{n\ell}+O(\varepsilon\mathcal{R}) & = & \frac{1}{12}\ddot{h}_{\ell\ell}+O(\varepsilon\mathcal{R}),\\
\Psi_{3} & = & -\frac{1}{2}\mathcal{R}_{n\ell n\overline{m}}+O(\varepsilon\mathcal{R}) & = & -\frac{1}{2}\overline{\mathcal{R}_{nm}}+O(\varepsilon\mathcal{R}) & = & \frac{1}{4}\ddot{h}_{\ell\overline{m}}+O(\varepsilon\mathcal{R}),\\
\Psi_{4} & = & -\mathcal{R}_{n\overline{m}n\overline{m}} & = & -\overline{\mathcal{R}_{nmnm}}+O(\varepsilon\mathcal{R}) & = & \frac{1}{2}\ddot{h}_{\overline{m}\overline{m}}+O(\varepsilon\mathcal{R}),\\
\Phi_{22} & = & -\mathcal{R}_{nmn\widetilde{m}} & = & -\frac{1}{2}\mathcal{R}_{nn}+O(\varepsilon\mathcal{R}) & = & \frac{1}{2}\ddot{h}_{m\overline{m}}+O(\varepsilon\mathcal{R}),
\end{array}\label{eq:POL_AMPL}
\end{equation}
\end{widetext}
which naturally are first order quantities.

Firstly, let us stress that in using the $E(2)$ framework,
in principle, there are 3 ways of calculating the polarization amplitudes:
directly calculating the Riemann tensor, the Ricci tensor or the individual
components of the metric perturbation. These can be viewed as the respective equivalences positions in Eq.\eqref{eq:POL_AMPL}. In our case, we consider the individual components of the metric perturbation. To further clarify the situation, we will give examples of GR and $f(\mathcal{R})$ in appendixes \ref{subsec:GR} and \ref{subsec:f(R)} respectively. In this setup, we will denote components of the linearized field equations, on a Minkowski background, by the symmetric tensor $\mathcal{H}_{\mu\nu}$, i.e $\mathcal{H}_{\mu\nu}=0$.

\section{Polarizations in GR\label{subsec:GR}}

In GR, the field equations are given by $G_{\mu\nu}=0,$ in vacuum, where $G_{\mu\nu}$ is the Einstein tensor. Linearizing these equations for a Minkowski background gives
\begin{equation}
    \mathcal{H}_{\mu\nu}\equiv\delta G_{\mu\nu}.
\end{equation}
Renaming and expressing in the N-P basis, the only non-trivial components read as
\begin{alignat}{2}
\mathcal{H}_{\mathit{m}\bar{\mathit{m}}} & =-\tfrac{1}{2}\mathit{\ddot{h}}_{\ell\ell} & \Rightarrow\Psi_{2}=0,\\
\mathcal{H}_{\mathit{n}\bar{\mathit{m}}}=\overline{\mathcal{H}_{\mathit{n}\mathit{m}}} & =-\tfrac{1}{2}\mathit{\ddot{h}}_{\ell\bar{\mathit{m}}} & \Rightarrow\Psi_{3}=0,\\
\mathcal{H}_{\mathit{n}\mathit{n}} & =-\mathit{\ddot{h}}_{\mathit{m}\bar{\mathit{m}}} & \Rightarrow\Phi_{22}=0.
\end{alignat}
Notice that $\Psi_{4}\neq0$ is unconstrained, and that  we have imposed a null wave. If one used an almost null
wave instead, it would turn out that all the polarization amplitudes are
zero, which means no wave at all. This is expected since we already
know that GR describes only a massless spin 2 particle. When the Ricci tensor is used to make this determination, the situation
is degenerate since $\delta\mathcal{R}_{\mu\nu}=0$ irrespective of what happens with the norm of the wave since in either case the only non trivial polarization amplitude is $\Psi_{4}$.

\section{Polarizations in \texorpdfstring{$f(\mathcal{R})$}{f(R)} Gravity\label{subsec:f(R)}}

In the same way, we study the case of $f(\mathcal{R})$, where the
background field equations, in vacuum, read as
\begin{equation}
f_{\mathcal{R}}\mathcal{R}_{\mu\nu}-\frac{1}{2}fg_{\mu\nu}+\left[g_{\mu\nu}\square-\mathcal{D}_{\mu}\mathcal{D}_{\nu}\right]f_{\mathcal{R}}=0,
\end{equation}
which after linearizing give
\begin{equation}
\mathcal{H}_{\mu\nu}\equiv\delta\mathcal{R}{}_{\mu\nu}-\tfrac{1}{2}g_{\mu\nu}\delta\mathcal{R}+C_{\mathcal{R}}\mathcal{R}{}_{\mu\nu}\delta\mathcal{R}=0,
\end{equation}
where it is assumed that $f_{\mathcal{R}}(0)\neq0$, $C_{\mathcal{R}}:=f_{\mathcal{R}\mathcal{R}}(0)/f_{\mathcal{R}}(0)$,
and again the calligraphic characters are quantities related to the
Levi-Civita connection. Expanding all terms in $h_{\mu\nu}$ and projecting in the N-P basis (assuming a null wave), we obtain the following non-trivial components for the linearized field equations
\begin{alignat}{2}
\mathcal{H}_{\mathit{m}\bar{\mathit{m}}} & =-\tfrac{1}{2}\mathit{\ddot{h}}_{\ell\ell} & \Rightarrow\Psi_{2}=0,\\
\mathcal{H}_{\mathit{n}\bar{\mathit{m}}}=\overline{\mathcal{H}_{\mathit{n}\mathit{m}}} & =-\tfrac{1}{2}\mathit{\ddot{h}}_{\ell\bar{\mathit{m}}} & \Rightarrow\Psi_{3}=0,\\
\mathcal{H}_{\mathit{n}\mathit{n}} & =-C_{\mathcal{R}}\ddddot{\mathit{h}_{\ell\ell}}-\mathit{\ddot{h}}_{\mathit{m}\bar{\mathit{m}}} & \;\;\;\Rightarrow\Phi_{22}\neq0.
\end{alignat}
Notice that the only difference between GR and $f(\mathcal{R})$
lies in the $\mathit{\ddot{h}}_{\mathit{m}\bar{\mathit{m}}}$ component
due to the presence of $C_{\mathcal{R}}\ddddot{\mathit{h}_{\ell\ell}}$
which is a pure $f(\mathcal{R})$ term since it is fourth order and
it is also multiplied by $C_{\mathcal{R}}$. Therefore $f(\mathcal{R})$
belongs to the $N_{3}$ class, if one assumes a null wave.

On the other hand, we also need to work through the same analysis by considering
an almost null wave, where $\eta_{\mu\nu}\widetilde{\ell}^{\mu}\widetilde{\ell}^{\nu}=\varepsilon<<1$,
for $f(\mathcal{R})$, since there is also a massive scalar mode present
\begin{widetext}
\begin{align}
\mathcal{H}_{\mathit{n}\mathit{n}} & =-C_{\mathcal{R}}\overset{....}{\mathit{h}_{\ell\ell}}-\mathit{\ddot{h}}_{\mathit{m}\bar{\mathit{m}}}+\varepsilon\Bigl(-\tfrac{1}{2}\mathit{\ddot{h}}_{\mathit{m}\bar{\mathit{m}}}+C_{\mathcal{R}}\bigl(-\overset{....}{\mathit{h}_{\ell\ell}}+3\overset{....}{\mathit{h}_{\ell\mathit{n}}}-2\overset{....}{\mathit{h}_{\mathit{m}\bar{\mathit{m}}}}\bigr)+\ddot{\mathit{h}}_{\mathit{n}\mathit{n}}\Bigr),\\
\mathcal{H}_{\mathit{n}\bar{\mathit{m}}}=\overline{\mathcal{H}_{\mathit{m}\mathit{n}}} & =-\tfrac{1}{2}\mathit{\ddot{h}}_{\ell\bar{\mathit{m}}}+\varepsilon(-\tfrac{1}{4}\mathit{\ddot{h}}_{\ell\bar{\mathit{m}}}+\tfrac{3}{4}\mathit{\ddot{h}}_{\mathit{n}\bar{\mathit{m}}}),\\
\mathcal{H}_{\mathit{m}\bar{\mathit{m}}} & =-\tfrac{1}{2}\ddot{\mathit{h}}_{\ell\ell}+\varepsilon(-\tfrac{1}{4}\mathit{\ddot{h}}_{\ell\ell}-C_{\mathcal{R}}\overset{....}{\mathit{h}_{\ell\ell}}+\tfrac{3}{2}\mathit{\ddot{h}}_{\ell\mathit{n}}-\tfrac{1}{2}\mathit{\ddot{h}}_{\mathit{m}\bar{\mathit{m}}}),\\
\mathcal{H}_{\ell\mathit{n}} & =\varepsilon\bigl(\tfrac{3}{2}C_{\mathcal{R}}\overset{....}{\mathit{h}_{\ell\ell}}-\mathit{\ddot{h}}_{\ell\mathit{n}}+\frac{3}{2}\mathit{\ddot{h}}_{\mathit{m}\bar{\mathit{m}}}\bigr),\\
\mathcal{H}_{\ell\ell} & =\varepsilon\mathit{\ddot{h}}_{\ell\ell},\\
\mathcal{H}_{\ell\bar{\mathit{m}}}=\overline{\mathcal{H}_{\ell\mathit{m}}} & =\tfrac{3}{4}\varepsilon\mathit{\ddot{h}}_{\ell\mathit{m}},\\
\mathcal{H}_{\bar{\mathit{m}}\bar{\mathit{m}}}=\overline{\mathcal{H}_{\mathit{m}\mathit{m}}} & =\tfrac{1}{2}\varepsilon\mathit{\ddot{h}}_{\bar{\mathit{m}}\bar{\mathit{m}}}
\end{align}
\end{widetext}
and having solved the system where the only nontrivial
polarization amplitude is $\Phi_{22}$ means that the theory
now belongs to the $O_{1}$ class. This is a consistent result since
if the theory did not really entail a massive scalar mode, then all
the polarization amplitudes would be zero, just as in GR above. 

Alternatively, using the approach where only the Ricci tensor is used in $f(\mathcal{R})$, as it is done throughout the literature \cite{Rizwana:2016qdq,Liang:2017ahj,Alves:2009eg}, one will find that the only non-trivial components $\mathcal{H}_{\mu\nu}\equiv\delta\mathcal{R}_{\mu\nu}$
are 
\begin{alignat}{2}
\mathcal{H}_{\mathit{n}\mathit{n}} & =-\tfrac{1}{6}\delta\mathcal{R}+C\delta\mathcal{R} & \Rightarrow\;\;\;\;\;\Phi_{22}\neq0,\\
\mathcal{H}_{\mathit{m}\bar{\mathit{m}}} & =\tfrac{1}{6}\delta\mathcal{R} & \Rightarrow\;inconclusive.
\end{alignat}
This means that we cannot use the Ricci scalar to determine the GW polarization properties for $f(Q,B)$. One can immediately see that $\Psi_{4}\neq0$ since there is no constraint
and also $\Phi_{22}\neq0$ while the rest are trivial. This is consistent with the previous results but apparently $\mathcal{H}_{\mathit{m}\bar{\mathit{m}}}\equiv\delta\mathcal{\mathcal{R}}_{\mathit{m}\bar{\mathit{m}}}\neq0$
is inconsistent and that's why the authors initially stated that the
brute force method of calculating the metric perturbation components
is the appropriate consistent approach.

\end{document}